\documentclass[onecolumn,notitlepage,amsmath,amssymb,aps,prd,longbibliography,nofootinbib,superscriptaddress,11pt]{revtex4-2}

\usepackage{hyperref}
\usepackage{url}
\usepackage{amsfonts,bm}
\usepackage[utf8]{inputenc}
\usepackage[T1]{fontenc}
\usepackage{hyperref}
\usepackage{graphicx,color}
\usepackage{bbold}
\usepackage{ulem}
\usepackage{wasysym}
\usepackage{verbatim}
\usepackage{epstopdf}
\usepackage{animate}
\usepackage{xmpmulti}
\usepackage{centernot}
\usepackage{multirow}
\usepackage{tikz}
\usepackage{pgfplots}
\pgfplotsset{compat=1.13}

\usepackage{graphicx}
\usepackage{float}
\usepackage{mathtools}
\usepackage{comment}
\usepackage{braket}

\usepackage{amsthm}
\theoremstyle{definition} 
\newtheorem{example}{Example}

\theoremstyle{theorem} 

\newcommand{\be}{\begin{equation}}
\newcommand{\ee}{\end{equation}}
\newcommand{\beq}{\begin{eqnarray}}
\newcommand{\eeq}{\end{eqnarray}}
\newcommand{\ba}{\begin{align}}
\newcommand{\ea}{\end{align}}

\newcommand{\up}{\uparrow}
\newcommand{\down}{\downarrow}

\def\C{\rm I\kern-.5em C} 
\def\co{\Delta}  

\def\be{\begin{equation}}
\def\ee{\end{equation}}
\def\bea{\begin{eqnarray}}
\def\eea{\end{eqnarray}}

\def\conm#1#2{\left [ {#1},{#2} \right ]}

\begin{document}
 \title{Deformations of the symmetric subspace of qubit chains}
 
 \author{Angel Ballesteros} 
   \email{angelb@ubu.es}
 \affiliation{Departamento de F\'isica, Universidad de Burgos, Pza. Misael Ba\~nuelos s.n., 09001 Burgos, Espa\~na}

 \author{Ivan Gutierrez-Sagredo} 
  \email{igsagredo@ubu.es}
 \affiliation{Departamento de Matem\'aticas y Computaci\'on, Universidad de Burgos, 
09001 Burgos, Spain}

 \author{Jose de Ram\'on \footnote{Corresponding author}} 
  \email{jderamon@ubu.es}
 \affiliation{Departamento de F\'isica, Universidad de Burgos, Pza. Misael Ba\~nuelos s.n., 09001 Burgos, Espa\~na}

\author{J. Javier Relancio} 
\email{jjrelancio@ubu.es}
 \affiliation{Departamento de Matem\'aticas y Computaci\'on, Universidad de Burgos, 
09001 Burgos, Spain}
  \affiliation{Centro de Astropart\'{\i}culas y F\'{\i}sica de Altas Energ\'{\i}as (CAPA),
Universidad de Zaragoza, Zaragoza 50009, Spain}

  \begin{abstract}
 The symmetric subspace of multi-qubit systems, that is, the space of states invariant under permutations, is commonly encountered in applications in the context of quantum information and communication theory.  It is known that the symmetric subspace can be described in terms of irreducible representations of the group $SU(2)$, whose representation spaces form a basis of symmetric states, the so-called Dicke states. In this work, we present deformations of the symmetric subspace as deformations of this group structure, which are promoted to a quantum group $\mathcal{U}_q(\mathfrak{su}(2))$. We see that deformations of the symmetric subspace obtained in this manner correspond to local deformations of the inner product of each spin, in such a way that departure from symmetry can be encoded in a position-dependent inner product. The consequences and possible extensions of these results are also discussed.   
\end{abstract}

\date{\today}
\maketitle


\bigskip
\bigskip

\vfill
\eject

\section{Introduction}
\label{sec:intro}

The symmetric subspace of multi-qubit systems, defined as the set of quantum states invariant under permutations \cite{Marcus1973,Marcus1975}, plays a fundamental role in quantum information theory, quantum computation, quantum optics, and quantum foundations \cite{watrous2018theory}. Symmetric pure states are an important resource in algorithms such as Grover's algorithm \cite{Grover1996}, {and the typically highly entangled nature of symmetric states~\cite{BG} also makes them crucial for applications in quantum metrology -where they are used to achieve Heisenberg-limited precision- or quantum networks -where they enable robust distribution of quantum entanglement- (see, for instance,~\cite{Barenco:1996kz,Zhang_2019,Campbell,Hansenne2022} and references therein).
} In the context of quantum optics they appear when considering certain collective phenomena in the light-matter interaction. In particular, in radiative processes it is predicted that a dense gas of atoms will decay by cascading down the symmetric subspace through a chain of highly entangled states known as Dicke states \cite{dicke1954coherence}, which form an orthonormal basis for the symmetric subspace. This process is the ultimate responsible for phenomena such as superradiance and subradiance \cite{Chumakov,Bienaime_2022,Gross_2022}. 

The symmetric subspace, and therefore Dicke states, can be constructed in terms of Lie group theory~\cite{ChumakovKlimov}. Indeed, given that the symmetric subspace can be thought of as an irreducible subrepresentation of the symmetric group of $N$ elements, $S_N$, in a set of $N$ qubits, it also admits a description in terms of the so-called \textit{Schur dual}~\cite{Dipper2008,weyl1939classical,Gross}, in this case irreducible representations of $SU(2)$.  

In this work, given the Lie theoretical description of the symmetric subspace, we analyze how its structure can be continuously deformed in the sense of Hopf algebra deformations of the universal enveloping algebra $\mathcal U(\mathfrak{su}(2))$ of $\mathfrak{su}(2)$. In particular, we deform the Lie algebra through the well-known $q$-deformation of the $\mathfrak{su}(2)$ Lie algebra, leading to the quantum algebra \( \mathcal{U}_q(\mathfrak{su}(2)) \). The resulting $q$-symmetric subspace retains many core features of Dicke states.  We investigate new features of the $q$-deformed symmetric subspace and its associated $q$-Dicke states \cite{zhang2009q,Raveh_2024}, {which have been recently shown to provide the ground states of the $\mathcal{U}_q(\mathfrak{su}(2))$ deformation of the Kittel--Shore model~\cite{qKSmodel}.} These $q$-Dicke states generalize the traditional Dicke states, forming a basis for the $q$-symmetric subspace while encoding deviations from symmetry through the deformation parameter \( q \). Importantly, we see that the $q$-deformation does not alter the local properties of individual qubits but rather modifies the global symmetry among qubits in the system. Furthermore, we show that the $q$-deformation can be interpreted as a localized modification of the Hilbert space inner product, providing a fresh perspective on how deformation parameters influence quantum states.

The $q$-representation, governed by the Hecke algebra \cite{chari1994guide}, captures a broader class of symmetric interactions and has potential implications for modeling real-world systems with imperfections or external perturbations. {In this context, we recall that the experimental preparation, control and detection of entanglement of Dicke states has achieved remarkable success by making use of different approaches and techniques (see, for instance, \cite{Retamal,Solano,kiesel2007experimental,wieczorek2009experimental,prevedel2009experimental,bartschi2019deterministic,mukherjee2020actual,aktar2022divide,bartschi2022short,li2021verification,raveh2024david})}. The generalized $q$-Dicke states, with their enriched structures {can thus be expected to provide new analytical approaches to -for instance- the description of certain fault-tolerant quantum computation protocols and enhanced quantum sensing proposals}. Moreover, these states may serve as a foundation for exploring the collective effects in systems with tunable symmetry, such as arrays of quantum dots, cold atoms, or photons in nonlinear optical cavities.  

 {This paper is organized as follows. Sections \ref{sec:symm_subsp} to \ref{sec:q-symm_subsp} introduce the notation, intuitions and main concepts of the rest of the paper. In particular, Section \ref{sec:symm_subsp} provides a foundational overview of the symmetric subspace and its representation-theoretical description. Section \ref{sec:su2} introduces the \( \mathcal U(\mathfrak{su}(2)) \) and \(\mathcal{U}_q(\mathfrak{su}(2)) \) Hopf algebras and the construction of Dicke states and $q$-Dicke states from the coproduct. This leads  to the definition of the $q$-symmetric subspace in Section \ref{sec:q-symm_subsp}. Sections \ref{sec:symm_qDicke} to \ref{sec:apl} contain the novel results of the paper. In particular, in Section \ref{sec:symm_qDicke}, we analyze the novel symmetries of $q$-Dicke states, whereas Section \ref{sec:q-rep} explores their $q$-representation in the symmetric group. Section \ref{sec:Hilbert} connects these findings to the deformation of the Hilbert space inner product. In Section \ref{sec:apl} potential applications of $q$-deformed symmetric subspaces of qubit chains to quantum informational and metrological tasks are sketched. Finally, Section \ref{sec:concl} outlines concluding remarks and directions for future research. }


\section{The symmetric subspace}
\label{sec:symm_subsp}
Consider a system whose state space is represented by $N$ tensor products of two-level systems or qubits, that is, its states are vectors in $\mathcal{H}=\bigotimes^N \mathcal{H}_{i}$, where $\mathcal{H}_{i}$ are identical two-dimensional Hilbert spaces.  Vectors in $\mathcal{H}$ can be represented as linear combinations of multiparty vectors of the form
\begin{align}\label{multi}
\ket{\psi}=\ket{\psi_1}\otimes...\otimes\ket{\psi_N},
\end{align}
with $\ket{\psi_i}\in \mathbb{C}^2$. 
In this work, we will focus on systems in which there is a total order relation in the qubits, that is, in which the qubits are ordered in a one-dimensional array. Hence, the system $\mathcal{H}$ should be envisioned as a {\it qubit chain}.  This structure, however, will not be relevant in the introductory sections, but eventually we will consider the deformations of certain subspaces of $\mathcal{H}$ that will depend explicitly on it. 

In this work, we shall investigate the subspace of permutation-invariant states and deformations thereof. Permutations in $\mathcal{H}$ are defined in the following way: For vectors of the form \eqref{multi}, one can perform permutations of the labels, 
\begin{align}
\ket{\tilde \psi}_\sigma=\ket{\psi_{\sigma(1)}}\otimes...\otimes\ket{\psi_{\sigma(N)}}, 
\end{align}
for any permutation $\sigma$ in the symmetric group of $N$ elements, $S_N$.  This operation for fixed $\sigma$ can be written, after extending this operation linearly to $\mathcal{H}$, as a linear operator of the form
\begin{align}
W(\sigma)\ket{\psi_1}\otimes...\otimes\ket{\psi_N}=\ket{\psi_{\sigma(1)}}\otimes...\otimes\ket{\psi_{\sigma(N)}}. 
\end{align}

It is obvious from the manner in which it acts on these vectors that $W(\sigma)W(\sigma')=W(\sigma\sigma')$, and in particular, $W(\sigma^{-1})=W^{-1}(\sigma)$. Therefore, $W$ is a linear representation of $S_N$ in $\mathcal{H}$.  In what follows, it will be useful to present the group $S_N$ as generated by the $N-1$ adjacent transpositions  {between the $i$-th and $i+1$-th element, which we denote} $t_i$ \cite{magnus2004group}. These adjacent transpositions fulfill the following relations
\begin{align}\label{tr}
    t_{i}t_{i+1}t_{i}&=t_{i+1}t_{i}t_{i+1},\\
     t_{i}t_{j}&= t_{j}t_{i}\quad |i-j|\geq 2,\\
     t_i^2&=\openone,
\end{align}
where $i,j \in \{1, \ldots,N-1\}$. The first two are the so-called braid relations, and $t_i^2=1$ reflects the fact that transpositions are idempotent. We will denote the representation on $\mathcal{H}$ of the adjacent transpositions $W_i:=W(t_i)$.

Moreover, given that 
\begin{align}
    \braket{\phi| W(\sigma)\psi}=\braket{\phi_1|\psi_{\sigma(1)}}...\braket{\phi_N|\psi_{\sigma(N)}}=\braket{\phi_{\sigma^{-1}(1)}|\psi_1}...\braket{\phi_{\sigma^{-1}(N)}|\psi_N},
\end{align}
we can conclude that $W(\sigma)^{\dagger}=W(\sigma^{-1})$. Hence, the representation is unitary.

Given that the group $S_N$ is finite, it is automatically semisimple \cite{weyl1939classical}. This just means that every representation space can be expanded in irreducible components, and in particular
\begin{align}\label{decomp1}
\mathcal{H}=\bigoplus_l \mathcal{H}_l\otimes M_l,
\end{align}
where $l$ runs over all the irreducible representations (irrep.) of $S_N$, and $\mathcal{H}_l$ are the corresponding representation spaces. The subspaces $M_l$ are the multiplicity subspaces associated with each irreducible representation. The subspace of $\mathcal{H}$ of permutation-invariant vectors, that is, vectors for which the representation $W(\sigma)$ acts trivially, is therefore the multiplicity space associated with the trivial representation, which is of course irreducible. This subspace will be henceforth called the {\it symmetric subspace}, and is denoted $\text{Sym} \, \mathcal{H}$.

 The decomposition in irreducible components of $\mathcal{H}$ in the case of the representation $W(\sigma)$ of $S_N$ is special. The multiplicity spaces for different $l$ values are all of different dimensions, and they correspond to irreducible representations of $GL(2)$. Hence, the subspaces $\mathcal{H}_l$ are the multiplicity subspaces of a distinguished representation of $GL(2)$. This representation is 
\begin{align}
    \rho^N_s(g)=g\otimes...\otimes g,
\end{align}
where $g\in GL(2)$ acts on $\mathcal{H}_{i}$ as a matrix in the defining representation. This important result is known as the Schur-Weyl duality \cite{Dipper2008,weyl1939classical}. The two groups that act on $\mathcal{H}$ and determine each other through their representations are said to be Schur dual. More concretely, the duality states that --with respect to the tensor product structure appearing in the decomposition \eqref{decomp1}--,
\begin{align}
    W(\sigma)=\bigoplus_l W_l(\sigma)\otimes \openone_l,
\end{align}
 where $W_l$ are all the unitary irreducible representations of $S_N$, labeled by Young tableaux, and
 \begin{align}
    \rho^N_s(g)=\bigoplus_l \openone_l\otimes \rho_l(g),
\end{align}
 where $\rho_l(g)$ are irreps of $GL(2)$.
 
 The symmetric subspace is given by the trivial subrepresentation of the symmetric group; therefore, it is also an irrep of $GL(2)$ according to the duality. In  particular, if one considers any symmetric vector $\psi$ in $\text{Sym}\mathcal{H}$ one can generate the whole symmetric subspace through $\rho^N_s$. Namely,
\begin{align}
\text{Sym} \, \mathcal{H}=\text{span }\;\rho^N_s(GL(2))\ket{\psi}.
\end{align}

 This construction holds for arbitrary representations of the permutation group on tensor products of vector spaces. However, each $\mathcal{H}_{i}$ is a Hilbert space, in addition to being a vector space, and the representation $W$ is unitary. Then one can show that actually 
\begin{align}
\text{Sym} \, \mathcal{H}=\text{span }\;\rho^N_s(U(2))\ket{\psi}.
\end{align}
where $U(2)$ denotes the unitary group in two dimensions. 
Even further, the representations of $U(2)$ appearing in the Schur-Weyl duality for $\mathcal{H}$ are not all possible representations of $U(2)$. For example, $\mathcal{H}$ does not carry the determinant representation $\pi(U)=|U|$, nor any dual representation. It can be shown \cite{Dipper2008} that this implies that every irreducible representation of $U(2)$ appearing in $\mathcal{H}$ is also an irreducible representation of $SU(2)$. 

\begin{example}
\label{ex:2q}
 Before proceeding, it is illustrative to analyze the simplest example of two qubits. Let $\{\ket{\up},\ket{\down}\}$ be a basis on the space of each individual qubit, $\mathcal{H}_{1}$, $\mathcal{H}_{2}$.  There is an orthonormal basis of the product of Hilbert spaces $\mathcal{H}=\mathcal{H}_1\otimes\mathcal{H}_2$ given by the vectors
\begin{align}
    \ket{\up}\ket{\up},\quad \ket{\down}\ket{\down},\quad\frac{\ket{\down}\ket{\up}+\ket{\up}\ket{\down}}{\sqrt 2},
\end{align}
and 
\begin{align}
   \frac{\ket{\down}\ket{\up}-\ket{\up}\ket{\down}}{\sqrt 2}.
\end{align}
The symmetric group for two elements, $S_2$, is generated by the identity and a single transposition $W$.
We observe that the first three vectors are invariant under $W$ and therefore, under all permutations.  Furthermore, the last flips a sign when it is acting with the elementary permutation $W$. This means that the subspace spanned by the first three vectors carries the trivial representation of $S_2$, and the subspace spanned by the last one the sign representation. Moreover, if we denote
\begin{align}
\ket{j=1,j_z=-1}=\ket{\down}\ket{\down},\;\ket{j=1,j_z=0}=\frac{\ket{\down}\ket{\up}+\ket{\up}\ket{\down}}{\sqrt 2},\;\ket{j=1,j_z=-1}=\ket{\up}\ket{\up},
\end{align}
and
\begin{align}
    \ket{j=0,j_z=0}=\frac{\ket{\down}\ket{\up}-\ket{\up}\ket{\down}}{\sqrt 2},
\end{align}
we see that these subspaces also carry irreducible representations of $SU(2)$ with $j=1,0$, respectively, which in fact are given by the decomposition in irreps of the tensor product of two $j=\frac12$ representations. We see that the irreducible representations of $S_2$ are thus paired with the irreducible representations of $SU(2)$.

\hfill $\square$
\end{example}




A fundamental feature of this duality is that the representations of $S_N$ are paired with representations of a simply connected compact group, which is in one-to-one correspondence with the universal enveloping algebra of a Lie algebra; in this case, $\mathfrak{su}(2)$. Universal enveloping algebras of Lie algebras are associative algebras complemented with a set of operations that turn them into Hopf algebras. These algebras, which we will introduce shortly, can be continuously deformed while preserving the structure of the representations. In the sequel we will describe the symmetric subspace in terms of the structure of Hopf algebra of the universal enveloping algebra of the Lie algebra $\mathfrak{su}(2)$, which will allow us to discuss deformations thereof.

\section{Hopf algebras and ($q-$)Dicke states}
\label{sec:su2}
A {Hopf algebra} is a (unital, associative) algebra $H$ endowed
with two homomorphisms called coproduct $(\Delta : H\to
H\otimes H )$ and counit $(\epsilon : H\to \mathbb{C})$, as well as an
antihomomorphism, called antipode $S : H\to H$), satisfying a set of compatibility conditions (see the standard texts \cite{majid1994foundations,chari1994guide,biedenharn1981tensor,kassel2012quantum,klimyk2012quantum} for more details). We remark that the whole Hopf algebra structure is instrumental for the results presented in the rest of the paper, although it is the coproduct map that will play the most remarkable role. 

In particular, the coassociativity of the coproduct map $\co$ allows the generalization to $N$ copies of $H$. If we denote $\co\equiv\co^{(2)}$,  the homomorphism $\co^{(N)}:H\rightarrow
H^{\otimes N}$ can be defined by the
recurrence relation
\be
\co^{(N)}:=(\co\otimes \text{id}\otimes\dots^{ {(N-2)}}\otimes \text{id})\circ\co^{(N-1)}.
\label{fla}
\ee

To discuss the aspects of unitarity, we will also need the concept of *-algebras. A *-structure in an associative algebra $\mathcal A$ is an antilinear anticomultiplicative map $*$ satisfying $*^2=\text{id}$. A Hopf algebra with a compatible *-structure is called a Hopf *-algebra \cite{majid1994foundations}.

 Group algebras and (the universal enveloping algebra of) Lie algebras can be endowed with these operations; in particular,  $\mathfrak{su}(2)$ can be described in these terms. Indeed, the universal enveloping algebra $\mathcal U (\mathfrak{g})$ of any Lie algebra $\mathfrak{g}$ is a Hopf algebra with coproduct given by
 \bea &\co(X_i)=\openone\otimes X_i + X_i\otimes \openone,\qquad
\co(\openone)=\openone\otimes \openone,\label{coprim}
\eea
where $X_i$ are the Lie algebra generators and $\openone$ denotes the identity in $\mathcal U (\mathfrak{g})$. These maps can be extended
to any monomial in $\mathcal U (\mathfrak{g})$ by means of the homomorphism condition for the coproduct 
$\co(XY)=\co(X)\co(Y)$, which implies the compatibility of the coproduct
$\Delta$ with the Lie bracket
\be
\conm{\co(X_i)}{\co(X_j)}_{H\otimes H}=\co(\conm{X_i}{X_j}_H),\qquad \forall
X_i,X_j\in \mathfrak{g}. \label{hom}
\ee
The coproduct can be used to generate higher dimensional representations from lower dimensional ones. Indeed, consider any representation of $\mathcal U (\mathfrak{g})$ in a Hilbert space $\mathcal{K}$, $\rho_{\mathcal{K}}$. We define tensor representations of the Hopf algebra through the generators $X_i$ as
\begin{align}
    \Delta^{(N)}_{\mathcal{K}}(X_i)&=\rho_{\mathcal{K}}(X_i)\otimes \openone\otimes
\dots^{ {(N-1)}}\otimes \openone \cr
& + \openone\otimes\rho_{\mathcal{K}} (X_i)\otimes \openone\otimes\dots^{ {(N-2)}}\otimes
 \openone +
\dots \cr
& + \openone\otimes
\dots^{ {(N-1)}}\otimes \rho_{\mathcal{K}}(X_i).
\label{coN}
\end{align} 
If a representation of $\mathcal U (\mathfrak{g})$ is irreducible, it maps the center of $\mathcal U (\mathfrak{g})$ to multiples of the identity because of Schur's lemma \cite{weyl1939classical}. When the center of the universal enveloping algebra is a finitely generated algebra, irreducible representations can be labeled using the proportionality constants of these generators. If $\lambda$ denotes such a set of labels determining an irrep, then we will write $\rho_\lambda$, rather than $\rho_{\mathcal{K}_\lambda}$, in order to lighten the notation. Similarly, we will denote tensor representations of irreps as $\Delta^{(N)}_{\lambda}$. Finally, if $\mathcal U(\mathfrak{g})$ is equipped with an involution *, the representation is said to be unitary if $\rho_{\mathcal{K}}(a^*)=\rho_{\mathcal{K}}(a)^\dagger$, where the dagger corresponds to the Hilbert space adjoint induced by the product in $\mathcal{K}$.  

\subsection{The $\mathcal{U}(\mathfrak{su} (2))$ Hopf algebra and Dicke states }

The $\mathfrak{su}(2)$ Lie algebra is defined by three generators with relations
\begin{equation}
    [J^3,J^\pm]= \pm J^\pm,\qquad [J^+,J^-]= 2 J^3.
    \label{eq:su2Lie}
\end{equation}
Its universal enveloping algebra, $\mathcal{U}(\mathfrak{su} (2))$, is a cocommutative Hopf algebra with coproduct given by
\begin{align}
&\Delta(J^3)=J^3\otimes\openone+\openone\otimes J^3, \qquad \Delta(J^\pm)=J^\pm\otimes\openone+\openone\otimes J^\pm,
\end{align}
Further,  with the involution defined by 
\begin{align}\label{invosu}
     (J^{3})^*=J^{3},\quad (J^{\pm})^*=J^{\mp},
\end{align}
the algebra $\mathcal U(\mathfrak{su}(2))$ becomes a Hopf *-algebra.

The unitary irreducible representations of $\mathcal U(\mathfrak{su}(2))$ are all finite-dimensional \cite{weyl1939classical}. In the case of $\mathcal U(\mathfrak{su}(2))$, the algebra generated by the Casimir element
\begin{align}
    \mathcal{C}=J^+J^{-}+J^3(J^3-\openone)
    \label{eq:cassu2}
\end{align}
is the full center of the Hopf algebra. Therefore, the representation of the Casimir and any function thereof is invariant and proportional to the identity for finite-dimensional irreducible representations. The proportionality constant labels different, inequivalent irreps.

If $k$ is the dimension of an irrep of $\mathcal U(\mathfrak{su}(2))$, then defining $j=\frac{k-1}{2}$, it follows that the Casimir element acts as
\begin{align}
    \rho_j(\mathcal{C})=j(j+1)\openone.
\end{align}

Further, we identify for this representation an orthogonal basis of $\mathcal{H}_j$ consisting of eigenvalues of $J^3$, $\{\ket{-j},...,\ket{j}\}$, such that
\begin{equation}
J^\pm |j, j_3\rangle\, :=\, \sqrt{(j\mp j_3) (j\pm j_3+1)}|j ,j_3\pm 1\rangle,\qquad  J^3 |j, j_3\rangle=j_3 |j, j_3\rangle.
\end{equation}
The fundamental $j=1/2$ irrep of $\mathfrak{su} (2)$ has thus a two dimensional representation space with basis
\begin{align}
    \ket{\frac12}=\ket{\up},\quad \ket{-\frac12}=\ket{\down},
\end{align}
which we call spin up and spin down, respectively.

This representation can be constructed in terms of the Pauli matrices 
\begin{equation}\label{actionqbits}
\sigma_x = 
\begin{pmatrix}
0 & 1 \\
1 & 0 \\
\end{pmatrix}, \qquad
\sigma_y = 
\begin{pmatrix}
0 & -i \\
i & 0 \\
\end{pmatrix}, \qquad
\sigma_z = 
\begin{pmatrix}
1 & 0 \\
0 & -1 \\
\end{pmatrix} \, ,
\end{equation}
such that
\begin{equation}
    \rho_{\frac12}(J^+) = \frac{\sigma_x+i\sigma_y}{2}=
\begin{pmatrix}
0 & 1 \\
0 & 0 \\
\end{pmatrix},
\qquad \rho_{\frac12}(J^-) = \frac{\sigma_x-i\sigma_y}{2}=
\begin{pmatrix}
0 & 0 \\
1 & 0 \\
\end{pmatrix} ,  \qquad \rho_{\frac12}(J^3) = \frac{1}{2} \sigma_z=\begin{pmatrix}
\frac12 & 0 \\
0 &  {-\frac12} \\
\end{pmatrix} .
\end{equation}

 Given this representation in a two-dimensional space, one can build a tensor representation $\Delta^{(N)}_{\frac12}$ on $N$ copies, that is a representation acting on $\mathcal{H}=\otimes^N \mathcal{H}_{\frac12}$.  Note that this representation is highly reducible 
 {(see~\cite{curtright2017spin} for the explicit computation of the multiplicities for each irreducible module)} and the task is to find the subspaces associated with its irreducible components.  One way to do so is to consider the eigenspaces of the Casimir operator and then build the representation spaces by acting with the reducible representation on an arbitrary element of the eigenspace.  For instance, one can consider the ``ground state'' with all spins down
\begin{align}
    \ket{G}:=\ket{\down...\down}.
\end{align}
The action of the representation over this vector forms an irreducible vector subspace for $\Delta^{(N)}_{\frac12}$ because this vector is an eigenvalue of the Casimir when represented through $\Delta^{(N)}_{\frac12}$. 
Indeed, since the Casimir element is given by \eqref{eq:cassu2} its representation is
\begin{align}
   \Delta^{(N)}_{\frac12} (\mathcal{C})=\Delta^{(N)}_{\frac12}(J^+) \Delta^{(N)}_{\frac12}(J^-) + \Delta^{(N)}_{\frac12}(J^3)\left(\Delta^{(N)}_{\frac12}(J^3)-\openone\right). 
\end{align}
However, $\Delta^{(N)}_{\frac12}(J^-)\ket{G}=0$ and $\ket{G}$ is an eigenvector  of $\Delta^{(N)}_{\frac12}(J^3)$, and it follows that $\ket{G}$ is an eigenvector of $ \Delta^{(N)}_{\frac12} (C)$. The associated eigenvalue is 
\begin{align}
     \Delta^{(N)}_{\frac12} (C) \ket{G}=\frac{N}{2}\left(\frac{N}{2}+1\right)\ket{G}.
\end{align}

Now, because the Casimir is in the center of $\mathcal{U}(\mathfrak{su}(2))$, its representation commutes with $ \Delta^{(N)}_{\frac12} (\mathcal{U}(\mathfrak{su}(2)))$. So
\begin{align}
   \Delta^{(N)}_{\frac12} (C) \;\Delta^{(N)}_{\frac12} (\mathcal{U}(\mathfrak{su}(2)))    \ket{G}=\frac{N}{2}\left(\frac{N}{2}+1\right)\Delta^{(N)}_{\frac12} (\mathcal{U}(\mathfrak{su}(2)))\ket{G}. 
\end{align}
We conclude that $\Delta^{(N)}_{\frac12} (\mathcal{U}(\mathfrak{su}(2)))\ket{G}$ is an irreducible representation space (also known as an irreducible module) of $\mathcal{U}(\mathfrak{su}(2))$, where $j=\frac{N}{2}$. The module is  {$(N+1)$} dimensional, with a basis of eigenvectors of $\Delta^{(N)}_{\frac12}(J^3)$ given by
\begin{align}
\ket{D^m_N}:=\mathcal{N}^N_m\left(\Delta^{(N)}_{\frac12}(J^+)\right)^m\ket{G},
\end{align}
for $m=j_3-N=0,...,N$. These states are known in the literature as {\it Dicke states}. The normalization constants can be calculated exactly,
\begin{equation}
    \mathcal{N}^N_m =\prod_{i=1}^m \frac{1}{\sqrt{(N+1-i) i}}= \frac{1}{m!} \frac{1}{\sqrt{\binom{N}{m}}}.
    \label{norDicke}
\end{equation}
Note that the factor $\frac{1}{m!}$ corrects for the consecutive application of the coproduct.

In the following examples we explicitly show how Dicke states for low $N=2,3$ are obtained from reiterated application of $\Delta_{\frac{1}{2}}(J^+)$.

\begin{example}
\label{ex:DickeN2}

For $N=2$, starting from $\ket{D_2^0} = \ket{\down \down}$, if we apply $\Delta^{(2)}_{\frac{1}{2}}(J^+)$ one time, we get
\begin{equation}
 \ket{D_2^1}= \frac{1}{\sqrt 2} \left( \Delta_{\frac12}^{(2)} (J_+) \right) \ket{D_2^0} =   \frac{1}{\sqrt 2} (\openone \otimes J_+ + J_+ \otimes \openone) \ket{\down \down} =  \frac{1}{\sqrt 2}(\ket{\down \up} + \ket{\up \down}).
\end{equation}
 Applying it one more time, we obtain
\begin{equation}
 \ket{D_2^2} =  \frac{1}{2}  \left( \Delta_{\frac12}^{(2)} (J_+) \right)^{2} \ket{D_2^0} = \ket{\up \up}.
\end{equation} 
Any further application of the coproduct yields the null vector, as should be. Recalling Example \ref{ex:2q}, we have that a basis of the symmetric subspace of a 2-qubit system is given by the Dicke states, \emph{i.e.} $\mathrm{Sym} \, \mathcal{H} = \text{span } \{ \ket{D_2^0}, \ket{D_2^1}, \ket{D_2^2} \}$.

\hfill $\square$
\end{example}

\begin{example}
\label{ex:DickeN3}

As in the previous example, for $N=3$, we have $\mathrm{Sym} \, \mathcal{H} = \langle \ket{D_3^0}, \ket{D_3^1}, \ket{D_3^2}, \ket{D_3^3} \rangle$, where the generated Dicke states are obtained from $\ket{D_3^0}$ by the reiterated application of $\Delta^{(3)}_{\frac{1}{2}}(J^+)$. Namely, we have
\begin{equation}
\ket{D_3^1}= \frac{1}{\sqrt 3} \left( \Delta_{\frac12}^{(3)} (J_+) \right) \ket{D_3^0}
= \frac{1}{\sqrt 3} \left( \ket{\down \down \up} + \ket{\down \up \down} + \ket{\up \down \down} \right) .
\end{equation}
 In the same manner,
\begin{equation}
\begin{split}
 \ket{D_3^2} = \frac{1}{2\sqrt 3} \left( \Delta_{\frac12}^{(3)} (J_+) \right)^2 \ket{D_3^0}
 =\frac{1}{\sqrt 3}   \left( \ket{\down \up \up} + \ket{\up \down \up} + \ket{\up \up \down} \right) ,
\end{split}
\end{equation}
and
\begin{equation}
\begin{split}
\ket{D_3^3} = \frac{1}{6} \left( \Delta_{\frac12}^{(3)} (J_+) \right)^3 \ket{D_3^0} = \ket{\up \up \up}.
\end{split}
\end{equation}

\hfill $\square$
\end{example}

Now, coming back to the original motivation, we would like to describe the symmetric representation of $U(2)$ in terms of tensor representations of $\mathcal{U}(\mathfrak{su}(2))$. Indeed, it turns out that \cite{weyl1939classical}
\begin{align}
    \Delta^{(N)}_{\frac12} (\mathcal{U}(\mathfrak{su}(2))) =\text{span }\rho^N_s(U(2)).
\end{align}
Given this result, we conclude that the irreducible component of $\Delta^{(N)}_{\frac12} (\mathcal{U}(\mathfrak{su}(2)))$ with $j=\frac{N}{2}$ is paired with the trivial representation of $S_N$ by Schur's duality. This means that the symmetric subspace is an $ {(N+1)}$ dimensional irreducible representation of $\mathcal{U}(\mathfrak{su}(2))$, 
\begin{align}
   \text{Sym} \, \mathcal{H}= \Delta^{(N)}_{\frac12} (\mathcal{U}(\mathfrak{su}(2)))\ket{G},
\end{align}
and also that the Dicke states form an orthonormal basis of the symmetric subspace of the multi-qubit system.

\subsection{The $\mathcal{U}_q(\mathfrak{su} (2))$ Hopf algebra}
\label{sec:suq2}

 {For $q\in \mathbb{C}$ not a root of unity}, the Hopf algebra $\mathcal{U}_q(\mathfrak{su}(2))$ is the algebra generated by elements $J^+,J^-,J^3$ modulo the relations
\begin{align}\label{qalgebra}
[J^3,J^\pm]=\pm J^\pm,\qquad [J^+,J^-] =[2J^3]_q,
\end{align}
where we have defined the $q$-number as
\begin{align}
    [x]_q=\frac{q^{\frac{x}{2}}-q^{-\frac{x}{2}}}{q^{\frac{1}{2}}-q^{-\frac{1}{2}}}.
\end{align}

Note that as $q\to1$, the defining relations become those of the Lie algebra $\mathfrak{su}(2)$, as $[x]_1=x$, and therefore by definition $\mathcal{U}_1(\mathfrak{su}(2))=\mathcal{U}(\mathfrak{su}(2))$. The deformed coproduct map is given by  
\begin{align}
&\Delta(J^3)=J^3\otimes\openone+\openone\otimes J^3, \qquad \Delta(J^\pm)=J^\pm \otimes q^{\frac{J^3}{2}}+q^{-\frac{J^3}{2}}\otimes J^\pm.
\label{eq:suqHopf}
\end{align}
{Notice that, for the sake of notational simplicity, the coproduct map for $\mathcal{U}_q(\mathfrak{su}(2))$ is denoted with the same symbols $\Delta$ as in the $\mathcal{U}(\mathfrak{su}(2))$ case. Also, in the $q$-deformed case the expression~\eqref{coN} for the $J^\pm$ generators transforms into
\begin{align}
    \Delta^{(N)}_{\mathcal{K}}(J^\pm)&=\rho_{\mathcal{K}}(J^\pm)\otimes \rho_{\mathcal{K}}(q^{\frac{J^3}{2}})\otimes
\dots^{ {(N-1)}}\otimes \rho_{\mathcal{K}}(q^{\frac{J^3}{2}}) \cr
& + \rho_{\mathcal{K}}(q^{-\frac{J^3}{2}})\otimes\rho_{\mathcal{K}} (J^\pm)\otimes (q^{\frac{J^3}{2}})\otimes\dots^{ {(N-2)}}\otimes
 (q^{\frac{J^3}{2}}) +
\dots \cr
& + \rho_{\mathcal{K}}(q^{-\frac{J^3}{2}})\otimes
\dots^{ {(N-1)}}\otimes \rho_{\mathcal{K}}(J^\pm)\, ,
\label{coNq}
\end{align}
where $\rho_{\mathcal{K}}$ a representation of $\mathcal{U}_q(\mathfrak{su}(2))$ on the Hilbert space denoted as $\mathcal{K}$. 
}

A characteristic feature of this type of deformation of the algebra, which holds whenever $q$ is not a root of unity, is that the representation theory of $\mathcal{U}_q (\mathfrak{su}(2))$ is just a smooth deformation of that for $\mathcal U (\mathfrak{su}(2))$. In particular, the unitary irreducible modules of $\mathcal{U}_q (\mathfrak{su}(2))$ are finite-dimensional, and every unitary finite-dimensional representation is completely reducible. Every irreducible module of $\mathcal U (\mathfrak{su}(2))$ is mapped continuously to an irreducible module  of $\mathcal{U}_q (\mathfrak{su}(2))$ with the same dimension as $q$ varies.
Moreover, the deformed Casimir operator for the algebra still generates the full center and labels the irreducible representations. Its expression is just given by
\begin{equation}
\mathcal{C}_q= J^- J^+ +[J^3]_q[J^3+\openone]_q=J^+ J^- + [J^3]_q[J^3-\openone]_q \, ,
\end{equation}
and its eigenvalues are $[j]_q[j+1]_q$ for $j$ half-integer. Again, for a $k$-dimensional irreducible module, $j$ is fixed and takes the value $j=\frac{k-1}{2}$. A basis of each irreducible module is given by eigenvalues of $J^3$, and the action of the generators in terms of this basis is deformed to
\begin{equation}
J^\pm |j, j_3\rangle\, :=\, \sqrt{[j\mp j_3]_q [j\pm j_3+1]_q}|j ,j_3\pm 1\rangle,\qquad  J^3 |j, j_3\rangle=j_3 |j, j_3\rangle\, ,
\end{equation}
where $j_3=-j,\dots, +j$.

The essential point for the results presented in this paper is the fact that the $j=1/2$ fundamental representation for $\mathcal U_q (\mathfrak{su}(2))$ obtained from~\eqref{qalgebra} coincides with the undeformed one given in~\eqref{eq:su2Lie}.

This implies that $\mathcal U_q (\mathfrak{su}(2))$ acts on a single qubit through \eqref{actionqbits}, and the effect of the $q$-deformation will arise when this action is generalized to multipartite states through the action of the $q$-deformed coproduct. This implies that this type of deformation does not change the internal structure of each qubit but only changes the notion of symmetry between qubits along the chain. In the following, we further investigate the impact of these deformations on permutation-invariant states.

\section{The \scalebox{1.4}{$q$}-symmetric subspace}
\label{sec:q-symm_subsp}

Given that deformations of the universal enveloping algebra of $\mathfrak{su}(2)$ for  {$q$ not a root of unity} preserve the structure of the representations, one could expect that the irreps contained in products of representations are in a way dual, in the Schur-Weyl sense, to irreducible representations of some deformation of the symmetric group. Indeed, this is the case, and is well known in the literature on quantum groups \cite{Gomez_Ruiz-Altaba_Sierra_1996,chari1994guide}. 

In short, the Hilbert space $\mathcal{H}$ can be decomposed into sectors as
\begin{align}
    \mathcal{H}=\bigoplus_l V_l\otimes \mathcal{H}_l
\end{align}
where $\mathcal{H}_l$ are distinct unitary irreducible representations of $\mathcal U_q (\mathfrak{su}(2))$. The $V_l$ are all distinct irreducible representations of an algebra, the so-called Hecke algebra. The Hecke algebra of $N$ elements $H_N$ is a finitely generated algebra with the following presentation
\begin{align}\label{hecke}
    A_{i}A_{i+1}A_{i}=A_{i+1}A_{i}A_{i+1},\\
     A_{i}A_{j}= A_{j}A_{i}\quad |i-j|\geq 2,\\
     A_i^2=\alpha A_i+\openone,
\end{align}
where $\alpha\in \mathbb{R}$.

We realize then that for $\alpha=0$ this is simply the presentation in terms of adjacent transpositions of the symmetric group \eqref{tr}, so the Hecke algebra is a deformation of the group algebra of the symmetric group. Let $R_i$ be the representation of the $R$-matrix in $V_i\otimes V_{i+1}$, and $W_i$ be the transposition between the corresponding tensor factors. Then, the Hecke algebra is represented in $\mathcal{H}$ by 
\begin{align}
    \pi[A_i]= q^{\frac{1}{4}}W_iR_i,
\end{align}
with $\alpha=q^{\frac{1}{2}}-q^{-\frac{1}{2}}$. This is a known result, so we will not discuss it further, but we refer the interested reader to \cite{Gomez_Ruiz-Altaba_Sierra_1996}. However, we mention that this is the most natural deformation of the symmetric group in the sense that one deforms the algebra of symmetric operators acting over $\mathcal{H}$. 
Similar to the case of the symmetric group, we define the $q$-symmetric subspace $\text{Sym}_q \, \mathcal{H}$ as the subspace carrying the trivial representation of the Hecke algebra. By arguments entirely analogous to the undeformed case, one can show that
\begin{align}
    \text{Sym}_q \, \mathcal{H}=\Delta^{(N)}_{\frac12} (\mathcal U_q(\mathfrak{su}(2)))\ket{G}, 
\label{eq:symqH}
\end{align}
where $\ket{G}$ is again the state with all the spins down (which is a Hecke invariant in addition to being permutation-invariant).

Being an irrep of $\mathcal{U}_q(\mathfrak{su}(2))$, the $q$-symmetric subspace admits a basis of vectors given by the repeated action of the representation of $J^+$. These are called the $q$-Dicke states \cite{Raveh_2024,zhang2009q} in analogy with the undeformed case, and {have the following expression in terms of the representation~\eqref{coNq} of the $N$-th coproduct map for the $\mathcal{U}_q(\mathfrak{su}(2))$ Hopf algebra}
\begin{align}
\ket{D^m_N}_q=\mathcal{N}^q_{N,m}\left( \Delta^{(N)}_{\frac{1}{2}}(J^+)\right)^m\ket{G},
\label{eq:qdickeG}
\end{align}
where $m=0,1,... N$, and 
\begin{align}
    \mathcal{N}^q_{N,m} = \frac{1}{[m]_q!} \frac{1}{\sqrt{\binom{N}{m}}_q},
\end{align}
and the $q$ binomial coefficients are given by
\begin{align}
    \binom{m}{r}_q=\frac{[m]_q!}{[m-r]_q![r]_q!}.
\end{align}

Therefore, the $q$-symmetric subspace is given by a $j=N/2$ representation of $\mathcal U_q(\mathfrak{su}(2))$, which is again  {$(N+1)$}-dimensional. 

 {We shall remark here that the explicit construction of arbitrary $q$-Dicke states given by \eqref{eq:qdickeG} is not the only possible formula for such states. In particular, an alternative version valid for qudit $q$-Dicke states, based on a purely combinatorial approach was presented in \cite{Raveh_2024}.}

We now explicitly consider the $q$-Dicke states corresponding to $N=2,3$ qubits. As is apparent from the expressions below, these $q$-Dicke states are deformations of the Dicke states from Examples \ref{ex:DickeN2} and \ref{ex:DickeN3}, where all the qubits are not similarly weighted.

\begin{example}
\label{ex:qDickeN2}

Following a procedure similar to that used in Example  \ref{ex:DickeN2}, we can construct all $q$-Dicke states for $N=2$ qubits starting from $\ket{G} = \ket{D^0_2}_q = \ket{\down \down}$ and applying repeatedly $\Delta(J^\pm)$ (see \eqref{eq:suqHopf}). In this way we have 
\begin{equation}
\ket{D^1_2}_q = \frac{1}{\sqrt{  {[2]_q}}} \left( \Delta_{\frac12}^{(2)} (J_+) \right) \ket{D_2^0}_q =   \frac{1}{\sqrt 2} (J^\pm \otimes q^{\frac{J^3}{2}}+q^{-\frac{J^3}{2}}\otimes J^\pm) \ket{\down \down} = 
\frac{1}{\sqrt{ {[2]_q}}}\left( q^{1/4} \ket{ \down\up} + q^{-1/4}\ket{\up\down} \right).
\end{equation}
 Applying it one more time, we obtain
\begin{equation}
 \ket{D_2^2}_q =  \frac{1}{ {[2]_q}}  \left( \Delta_{\frac12}^{(2)} (J_+) \right)^{2} \ket{D_2^0}_q = \ket{\up \up}.
\end{equation} 
Further applying the coproduct results in the null vector. Therefore, from \eqref{eq:symqH}, we have $\mathrm{Sym}_q \, \mathcal{H} = \text{span } \{ \ket{D_2^0}_q, \ket{D_2^1}_q, \ket{D_2^2}_q \}$.

\hfill $\square$
\end{example}

\begin{example}
\label{ex:qDickeN3}

Similarly, for $N=3$, by applying $\Delta(J^\pm)$ to $\ket{G} = \ket{D^0_3}_q = \ket{\down \down \down}$ we construct all $q$-Dicke states, which are a $q$-deformation of the ones in Example \ref{ex:DickeN2}. We obtain
\begin{align}
 \ket{D^1_3 }_q&=\mathcal N_{3,1}^q\,\Delta^{(3)} (J^+) \ket{G}=\frac{1}{\sqrt{ {[3]_q}}}\left( q^{1/2}|\down\down\up\rangle +|\down\up\down\rangle+ q^{-1/2}|\up\down\down\rangle\right), \\
\ket{D^2_3 }_q&=\mathcal N_{3,2}^q\,[\Delta^{(3)} (J^+)]^2 \ket{G}=\frac{1}{\sqrt{ {[3]_q}}}\left( q^{1/2} |\down\up\up\rangle +|\up\down\up\rangle + q^{-1/2}|\up\up\down\rangle \right), \\
 \ket{D^3_3 }_q&= \mathcal N_{3,3}^q\,[\Delta^{(3)} (J^+)]^3 \ket{G}= \ket{\up\up\up} ,
\label{eq:3q3/2}
\end{align}
and thus $\mathrm{Sym}_q \, \mathcal{H} = \text{span } \{ \ket{D_3^0}_q, \ket{D_3^1}_q, \ket{D_3^2}_q, \ket{D_3^3}_q \}$.

\hfill $\square$
\end{example}

\section{A new symmetry of the $q$-Dicke states}
\label{sec:symm_qDicke}

The representations of the Hecke algebra, which have been extensively studied (\cite{Gomez_Ruiz-Altaba_Sierra_1996} and references therein), perhaps complicate too much the representation theory when it comes just to the characterization of the $q$-symmetric subspace. While the Hecke algebra characterizes the full deformed structure of the symmetric group in $\mathcal{H}$, it would be desirable to find a characterization that is unique to the trivial representation and that can be easily encoded in representations of the symmetric group. In what follows, we show that this is the case by closely examining novel properties of the $q$-Dicke states.
Now, we present a symmetry of the $q$-Dicke states that, to our knowledge, is not directly related to the $R$-matrix, in the sense that it cannot be derived straightforwardly from the defining properties of quasitriangular Hopf algebras. 

To understand this symmetry, we first describe it in the simplest case, the product representation of two identical fundamental irreps of $\mathcal{U}_q(\mathfrak{su}(2))$ with irrep label $j=\frac{1}{2}$. The irrep component in $\mathcal{H}$ with $j=1$ fulfills $\ket{j=1, j_3=-1}=\ket{G}=\ket{\down \down}$, and the remaining $q$-Dicke states
\begin{align}
    \ket{D^m_2}_q= \mathcal{N}^q_{2,m}\left(\Delta_{\frac{1}{2}}(J^+)\right)^m\ket{\down \down},
\end{align}
for $m=0,1,2$, span the $q$-symmetric subspace. These states have additional structure. First, note that substituting $q$ by $q^{-1}$
inverts the factors in the coproduct for all generators. This means in particular that
\begin{align}
    W\ket{D^m_2}_q= \mathcal{N}^q_{2,m}\left(\bar{\Delta}_{\frac{1}{2}}(J^+)\right)^m W\ket{\down\down}=\mathcal{N}^q_{2,m}\left(\bar{\Delta}_{\frac{1}{2}}(J^+)\right)^m \ket{\down\down}=\ket{D^m_2}_{q^{-1}}
\end{align}
where $W$ is the elementary transposition, and we have used that the normalization constants $\mathcal{N}^q_{2,m}$ are rational functions of $q$-numbers that are invariant under the change $q\to q^{-1}$.
The only  {$q$}-Dicke state that is not invariant under $W$ (that is, that depends explicitly on the parameter $q$) is $\ket{D^1_2}_q$, which again is given by
\begin{equation}
\ket{D^1_2}_q = 
\frac{1}{\sqrt{ {[2]_q}}}\left( q^{1/4} \ket{ \down\up} + q^{-1/4}\ket{\up\down} \right).
\end{equation}
It is not difficult to see that transformation $q\to q^{-1}$ can be implemented through a product operator: 
\begin{align}
    \ket{D^1_2}_q=\frac{\mathcal{N}^q_{2,1}}{\mathcal{N}^{q^{-1}}_{2,1}}q^{-\frac12J^3}\otimes q^{\frac12J^3}\ket{D^1_2}_{q^{-1}} =q^{-\frac12J^3}\otimes q^{\frac12J^3}W\ket{D^1_2}_{q},
\end{align}
where we have used that the normalization is invariant under the change $q\to q^{-1}$. In fact, given that $\ket{D^0_2}_q$ and $\ket{D^2_2}_q$ are also (rather trivially) invariant under this transformation, we conclude that for $N=2$ the operator defined as 
\begin{align}
    W^q=q^{-\frac{J^3}{2}}\otimes q^{\frac{J^3}{2}}W
\label{eq:WqN2}
\end{align}
fulfills
\begin{align}
   W^q \ket{D^m_2}_q=\ket{D^m_2}_q
\label{eq:WqD2}
\end{align}
for all $m$. As the $q$-Dicke states span the $q$-symmetric subspace, $W^q$ acts trivially over it, so it determines a symmetry of this subspace. Next, we will show that the coassociativity of the coproduct implies that  this simple invariance can be promoted to higher number representations. Indeed, consider an arbitrary $q$-Dicke state of $N$ qubits $\ket{D^m_N}_q$.  Its expression in terms of the coproduct is
\begin{align}
    \ket{D^m_N}_q=\mathcal{N}^q_{N,m}\left(\Delta^{(N)}_{\frac{1}{2}}(J^+)\right)^m\ket{\down ... \down}=\Delta^{(N)}_{\frac{1}{2}}(h)\ket{\down ... \down},
\end{align}
where we have defined $h=\mathcal{N}^q_{2,m}(J^+)^m$ and used the fact that $\Delta^{(N)}_{\frac{1}{2}}$ is a linear homomorphism. 
{Now, if we consider the $h$ operator for the site $i$, namely $h_{(i)}=\mathcal{N}^q_{2,m}(J_{(i)}^+)^m$,
due to the coassociativity property of the coproduct map $\Delta$ 
\begin{align}
    \Delta^{(N)}=(\text{id}\otimes\dots \otimes\Delta_i\otimes \dots\otimes \text{id})\circ \Delta^{(N-1)},
\end{align}
where $\Delta_i$ maps the $i$-th copy of $\mathcal{U}_q(\mathfrak{su}(2))$ into the tensor product of the $(i,i+1)$ copies. For the chosen representation this implies
\begin{align}
    \Delta^{(N)}_{\frac{1}{2}}(h)=(\text{id}\otimes\dots\otimes (\Delta_{\frac{1}{2}})_i\otimes \dots\otimes \text{id})\circ \Delta^{(N-1)}_{\frac{1}{2}}(h),
\end{align}
and by making use of Sweedler's notation $\Delta^{(N-1)}_{\frac{1}{2}}(h)=\sum h_{(1)}\otimes...\otimes h_{(i)}\otimes \dots \otimes h_{(N-1)}$ we can thus write
\begin{align}
    \Delta^{(N)}_{\frac{1}{2}}(h)=\sum h_{(1)}\otimes...\otimes\Delta_{\frac{1}{2}}(h_{(i)})\otimes...\otimes h_{(N-1)}.
\end{align}
Therefore, the $q$-Dicke states can be written as
\begin{align}
\ket{D^m_N}_q=\sum h_{(1)}\ket{\down}\otimes...\otimes\left(\Delta_{\frac{1}{2}}(h_{(i)})\ket{\down\down}\right)\otimes...\otimes h_{(N-1)})\ket{\down}.
\end{align}
}

We realize that the Dicke states can be written as the sum of states whose $(i,i+1)$ tensor component is of the form $\Delta_{\frac{1}{2}}(h_{(i)})\ket{\down \down}$. Therefore, this component belongs to the $q$-symmetric subspace of two qubits. Now, define
    \begin{align}
    W_i^q=q^{-\frac{J^3_i}{2}}\otimes q^{\frac{J^3_{i+1}}{2}}W_i,
\end{align}
where $W_i$ is again the adjacent transposition between the $i-$th and $(i+1)-$th sites. Because the $(i,i+1)$ tensor component of any Dicke state is $q$-symmetric, we find that 
\begin{align}
    \ket{D^m_N}_q=W^{q}_i\ket{D^m_N}_q
\label{eq:WqDN}
\end{align}
for all $i$. Finally, we realize that because the $q$-Dicke states span the $q$-symmetric subspace, all the $W^{q}_i$ act trivially on it. We shall call this set of $(N-1)$ operators $W^{q}_i$  as {\it q-transpositions}.

In the following examples, we discuss explicitly the cases of $N=2$ and $N=3$ qubits considered in Examples \ref{ex:qDickeN2} and \ref{ex:qDickeN3}.

\begin{example}
\label{ex:WqN2}
Identifying $\mathcal H_{\frac{1}{2}} \approx \mathbb{C}^2$ and setting 
\begin{equation}
    \ket{\up} = \begin{pmatrix}
        1 \\
        0
    \end{pmatrix}, \qquad 
    \ket{\down} = \begin{pmatrix}
        0 \\
        1
    \end{pmatrix},
\end{equation}
we have that
\begin{equation}
    \ket{D_2^0}_q = \begin{pmatrix}
        0 \\
        0 \\
        0 \\
        1 \\
    \end{pmatrix}, \qquad \qquad
    \ket{D_2^1}_q = \frac{1}{\sqrt{ {[2]_q}}} \begin{pmatrix}
        0 \\
        q^{-\frac{1}{4}} \\
        q^{\frac{1}{4}} \\
        0 \\
    \end{pmatrix}, \qquad \qquad
    \ket{D_2^2}_q = \begin{pmatrix}
        1 \\
        0 \\
        0 \\
        0 \\
    \end{pmatrix}. 
\end{equation}
The only $q$-transposition matrix in this case, given by \eqref{eq:WqN2}, explicitly reads
\begin{equation}
W^q_1=
\left(
\begin{array}{cccc}
 1 & 0 & 0 & 0 \\
 0 & 0 & q^{-\frac{1}{2}} & 0 \\
 0 & q^{\frac{1}{2}} & 0 & 0 \\
 0 & 0 & 0 & 1 \\
\end{array}
\right) ,
\end{equation}
from where we recover \eqref{eq:WqD2} in matrix form. Note that in the limit $q \to 1$ we recover the usual permutation matrix.

\hfill $\square$
\end{example}

\begin{example}
\label{ex:WqN3}
For $N=3$ qubits, the $q$-Dicke states from Example \ref{ex:qDickeN3} read
\begin{equation}
    \ket{D_3^0}_q = \begin{pmatrix}
        0 \\
        0 \\
        0 \\
        0 \\
        0 \\
        0 \\
        0 \\
        1 \\
    \end{pmatrix}, \qquad \quad
     \ket{D_3^1}_q = \frac{1}{\sqrt{ {[3]_q}}}\begin{pmatrix}
        0 \\
        0 \\
        0 \\
        q^{-\frac{1}{2}} \\
        0 \\
        1 \\
        q^{\frac{1}{2}} \\
        0 \\
    \end{pmatrix}, \qquad \quad
    \ket{D_3^2}_q = \frac{1}{\sqrt{ {[3]_q}}}\begin{pmatrix}
        0 \\
        q^{-\frac{1}{2}} \\
        1 \\
        0 \\
        q^{\frac{1}{2}} \\
        0 \\
        0 \\
        0 \\
    \end{pmatrix}, \qquad \quad
    \ket{D_3^3}_q = \begin{pmatrix}
        1 \\
        0 \\
        0 \\
        0 \\
        0 \\
        0 \\
        0 \\
        0 \\
    \end{pmatrix}.
\end{equation}
In this case we have two $q$-transpositions, whose matrices are given by
\begin{equation}
    W^q_{1} = 
\left(
\begin{array}{cccccccc}
 1 & 0 & 0 & 0 & 0 & 0 & 0 & 0 \\
 0 & 1 & 0 & 0 & 0 & 0 & 0 & 0 \\
 0 & 0 & 0 & 0 & q^{-\frac{1}{2}} & 0 & 0 & 0 \\
 0 & 0 & 0 & 0 & 0 & q^{-\frac{1}{2}} & 0 & 0 \\
 0 & 0 & q^{\frac{1}{2}} & 0 & 0 & 0 & 0 & 0 \\
 0 & 0 & 0 & q^{\frac{1}{2}} & 0 & 0 & 0 & 0 \\
 0 & 0 & 0 & 0 & 0 & 0 & 1 & 0 \\
 0 & 0 & 0 & 0 & 0 & 0 & 0 & 1 \\
\end{array}
\right), \qquad \qquad
    W^q_{2} = 
\left(
\begin{array}{cccccccc}
 1 & 0 & 0 & 0 & 0 & 0 & 0 & 0 \\
 0 & 0 & q^{-\frac{1}{2}} & 0 & 0 & 0 & 0 & 0 \\
 0 & q^{\frac{1}{2}} & 0 & 0 & 0 & 0 & 0 & 0 \\
 0 & 0 & 0 & 1 & 0 & 0 & 0 & 0 \\
 0 & 0 & 0 & 0 & 1 & 0 & 0 & 0 \\
 0 & 0 & 0 & 0 & 0 & 0 & q^{-\frac{1}{2}} & 0 \\
 0 & 0 & 0 & 0 & 0 & q^{\frac{1}{2}} & 0 & 0 \\
 0 & 0 & 0 & 0 & 0 & 0 & 0 & 1 \\
\end{array}
\right) .
\end{equation}
Again, form this expressions it is apparent that \eqref{eq:WqDN} holds, and in the limit $q \to 1$ we recover the permutation matrices that play the role of symmetries of the usual Dicke states.

\hfill $\square$
\end{example}

Hence, we found a new symmetry of the $q$-symmetric subspace. In what follows, we will show that this symmetry forms a representation of the symmetric group, although, as we will see, it is not unitary. 

\section{A  \scalebox{1.4}{$q$}-representation of the symmetric group}
\label{sec:q-rep}

We have found that the set of operators $W^q_i$ acts trivially over the $q$-symmetric subspace. Therefore, the algebra generated by all of them also acts trivially. How can we characterize this algebra? We will show that the algebra generated by the $W^q_i$ is actually a representation of the symmetric group algebra. 

We begin by introducing some notation, and we define the operators 
\begin{align}
    C^q_{i}=q^{-\frac{J^3_i}{2}}\otimes q^{\frac{J^3_{i+1}}{2}},
\end{align}
in such a way that
\begin{align}\label{qtrans}
    W^{q}_i=C^q_{i}W_i.
\end{align}
The operators $C^q_{i}$ have the following properties for $q>0$:
\begin{align}
(C^q_{i})^{\dagger}=C^q_{i},\qquad (C^q_{i})^{-1}=C^{q^{-1}}_{i}=W_iC^{q}_{i}W_i.
\end{align}

We want to characterize the algebra generated by all the $q$-transpositions. In order to do so, we analyze arbitrary products of the form
\begin{align}
    \prod^{\leftarrow} W^q_{a_i}=W^q_{a_M}...W^q_{a_1},
\end{align}
where $a_i$ is a sequence of $M$ elements that take values in $1,2,...,N-1$.

\begin{figure}[!htb]
    \centering
    \begin{minipage}{.5\textwidth}
        \centering
        \includegraphics[width=1\linewidth, height=0.15\textheight]{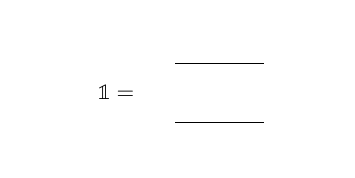}
        \caption{Identity diagram}
        \label{fig:identity}
    \end{minipage}%
    \begin{minipage}{0.5\textwidth}
        \centering
        \includegraphics[width=1\linewidth, height=0.15\textheight]{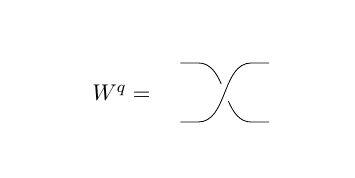}
        \caption{Elementary transposition}
        \label{fig:qtrans}
    \end{minipage}
\end{figure}

Although the analysis, of course, can be performed algebraically, perhaps it is more intuitive to represent such products through diagrams. For any product of $q$-transpositions, we associate a horizontal diagram of $N$ strands $\mathcal{D}$ in which each factor in the (ordered) product of $q$-transpositions is sequentially represented by a crossing in which the lower strand crosses on top of the upper strand. For example, for two qubits, the identity is associated with the diagram~\ref{fig:identity} and the elementary transposition with the diagram~\ref{fig:qtrans}.

 If we consider the action of product of $q$-transpositions associated with a particular diagram $\mathcal{D}$, which we momentarily denote $W^q[\mathcal{D}]$, on a product state $\ket{\Psi}=\ket{\psi_1}\otimes...\otimes\ket{\psi_N}$, it is not difficult to see that
\begin{align}
W^q[\mathcal{D}]\ket{\Psi}=O_{\mathcal{D},1}\ket{\psi_{\sigma(1)}}\otimes...\otimes O_{\mathcal{D},N}\ket{\psi_{\sigma(N)}} { ,}
\end{align}
where $\sigma$ is the permutation taking the $N$ elements from their initial position in the diagram $\mathcal{D}$ to their final position, and 
\begin{align}\label{crossings}
    O_{\mathcal{D},i}= q^{\frac{J^3}{2}M_{\mathcal{D},i}},
\end{align}
where we have defined the integer $M_{\mathcal{D},i}$ as the number of upper-crossings minus the number of lower-crossings in the path from $\sigma^{-1}(i)$ to $i$ in $\mathcal{D}$.

\begin{figure}[h!]
    \centering
    \includegraphics[width=0.5\linewidth]{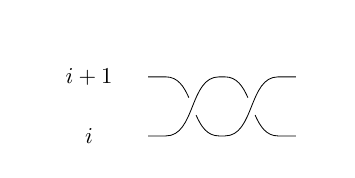}
    \caption{Squared $q$-transposition}
    \label{fig:Sqtrans}
\end{figure}

Because product states span $\mathcal{H}$, we conclude that any product of $q$-transpositions can be written as
\begin{align}\label{diagra}
    W^q[\mathcal{D}]=\left(\bigotimes^N_{i=1}O_{\mathcal{D},i} \right)W(\sigma).
\end{align}
In this notation it is straightforward to check that
\begin{align}\label{qidem}
    (W^q_i)^2=\openone, 
\end{align}
since this product is associated with the diagram~\ref{fig:Sqtrans}.
We see that each strand has both a lower-crossing and an upper-crossing, and both recover their initial positions. Therefore, according to Equation \eqref{diagra}, this is the identity operator.

\begin{figure}[h!]
    \centering
    \includegraphics[width=0.45\linewidth]{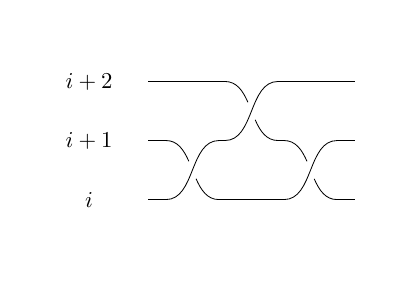}
    \includegraphics[width=0.45\linewidth]{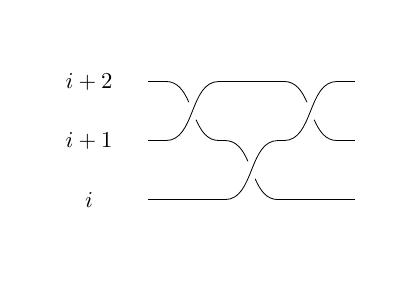}
    \caption{Braid relations}
    \label{fig:braid}
\end{figure}

Further, it is also evident that the $q$-transpositions satisfy the braid property
\begin{align}\label{qbride}
    W^q_iW^q_{i+1}W^q_i=W^q_{i+1}W^q_iW^q_{i+1}
\end{align}
because the diagrams~\ref{fig:braid}
induce the same permutation, and also the number of upper and lower-crossings is identical for each strand. Therefore, the algebra generated by the $q$-transpositions fulfills the algebraic relations of the adjacent transpositions in \eqref{tr}.

This means that the algebra generated by the $q$-transpositions is actually a representation of the symmetric group of $N$ elements. The representation is defined by first decomposing the permutation as a product of adjacent transpositions and then constructing the diagram $\mathcal{D}$ associated with such decomposition.  We then represent the permutation in $\mathcal{H}$ as 
\begin{align}
    W^q(\sigma)=C(\sigma) W(\sigma),
    \label{qper}\end{align}
where 
\begin{align}
C(\sigma)=\bigotimes^N_{i=1}C_{\mathcal{D},i}. 
\end{align}

In summary, we have found a representation $W^q$ of $S_N$ that leaves $\text{Sym}_q\mathcal{H}$ invariant. In fact, $\text{Sym}_q\mathcal{H}$ is completely characterized by this representation. However, this representation is not unitary. Indeed, the $q$-transpositions fulfill
\begin{align}\label{adjtrans}
  (  W^{q}_i)^\dagger=(W_i)^\dagger(C^q_{i})^\dagger=W_iC^q_{i}=C^{q^{-1}}_{i}W_i=W^{q^{-1}}_i.
\end{align}
Therefore, the representation is not unitary unless $q=1$, which corresponds to an ordinary representation of the symmetric group. However, there is a simple relationship between the adjoint and the inverse:
\begin{align}
    (  W^{q}(\sigma))^\dagger=W^{q^{-1}}(\sigma^{-1}),
\end{align}
which can be seen by decomposing an arbitrary $q$-permutation into $q$-transpositions and applying \eqref{adjtrans}.

 Many interesting properties of the undeformed representation stem from the fact that it is unitary; therefore, it would be desirable to map the $q$-representation to a unitary representation. We can do so by defining a deformed conjugation $*$ such that $(W^{q}_i)^*=W^{q}_i$, and extending it to the entire symmetric group such that $(  W^{q}(\sigma))^*=W^{q}(\sigma^{-1})$.
 The new conjugation can be defined by first considering the order-reversing permutation $\tau$, which is defined as the permutation that acts as $\tau(i)=N+1-i$. It is clear that $\tau^2=1$, and that for any adjacent transposition $\tau t_i \tau= t_{N-i}$.

\begin{figure}[h!]
    \centering
    \includegraphics[width=\linewidth]{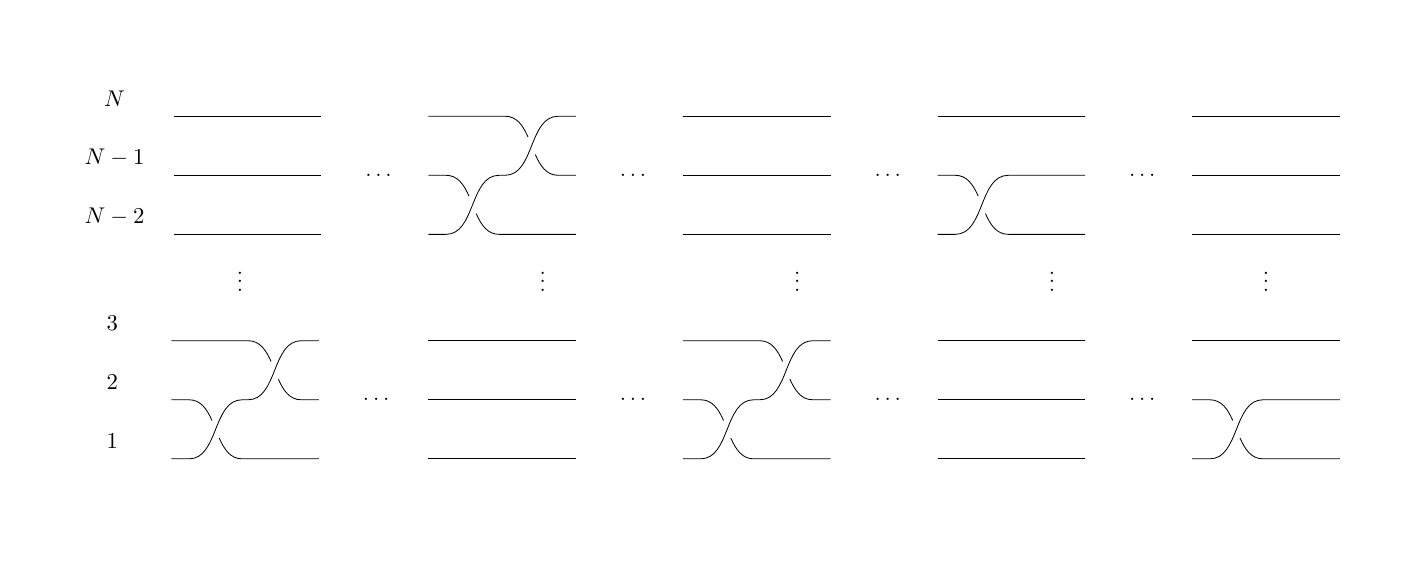}
    \caption{Total inversion in terms of adjacent transpositions}
    \label{fig:trenzagrande}
\end{figure}

Consider the adjoint action of the $q$-representation of $\tau$ acting on an arbitrary $q$-transposition. It obviously holds that
\begin{align}
    W^{q}(\tau)W^{q}_iW^{q}(\tau)=W^{q}_{N-i},
\end{align}
 because $W^{q}(\sigma)$ is a representation of $S_N$. On the other hand, the action of the ordinary order reversing on the $q$-transpositions is
\begin{align}
    W(\tau)W^{q}_iW(\tau)=W^{q^{-1}}_{N-i},
\end{align}
which can be easily obtained from Equation \eqref{qtrans}. We now concatenate these transformations:
\begin{align}
    W(\tau)W^{q}(\tau)W^{q}_iW^{q}(\tau)W(\tau)=W^{q^{-1}}_{i}=(  W^{q}_i)^\dagger,
\end{align}
from which we finally conclude that
\begin{align}
    W^{q}(\tau)W(\tau)(  W^{q}_i)^\dagger W(\tau)W^{q}(\tau)=W^{q}_{i}.
\end{align}
Now, we recall Equation \eqref{qper}, from which we obtain the following expression:
\begin{align}
    W^{q}(\tau)W(\tau)=C(\tau)(W(\tau))^2=C(\tau),
\end{align}
and also
\begin{align}
   W(\tau) W^{q}(\tau)=C^{-1}(\tau).
\end{align}
In summary,
\begin{align}
    C(\tau)(  W^{q}_i)^\dagger C^{-1}(\tau)=W^{q}_{i}
\end{align}
for all adjacent transpositions. Because the operation $C(\tau)\cdot C^{-1}(\tau)$ is a homomorphism, it extends to the whole representation, from which we conclude that
\begin{align}
    C(\tau)(  W^{q}(\sigma))^\dagger C^{-1}(\tau)=W^{q}(\sigma^{-1}).
\end{align}
The expression for $C(\tau)$ is relatively easy to obtain by factorizing $\tau$ as the following product of adjacent transpositions
\begin{align}
    \tau=t_1t_2t_1...\left(\prod^{N-2}_{i=1}t_i\right)\left(\prod^{N-1}_{i=1}t_i\right)=\prod^{N-1}_{k=1}\left(\prod^{N-k}_{i=1}t_i\right),
\end{align}
corresponding to the diagram \ref{fig:trenzagrande}.

From this diagram, paying attention to the crossings, it can be seen from inspection that the number of upper-crossings of the $i$-th strand is $i-1$, and the number of lower-crossings is $N-i$. Hence, using Equation \eqref{crossings} we obtain
\begin{align}
    C(\tau)=q^{-\frac12\sum^{N}_{i=1} (N+1-2i)J_i^3}.
\end{align}

\section{Deformation of the symmetric subspace as a deformation of the Hilbert space's product}
\label{sec:Hilbert}

The last section analyzes the structure of a non-unitary representation of the symmetric group with the $q$-symmetric subspace as its invariant subspace. Recall that we have found an operation, such that
\begin{align}
    W^{q}(\sigma^{-1})=C(\tau)(  W^{q}(\sigma))^\dagger C^{-1}(\tau)
\end{align}
for all permutations. 

This means that by defining an operation $*$ in $B[\mathcal{H}]$, the bounded operators acting on $\mathcal{H}$,  given by
\begin{align}\label{invo}
    A^*:=C(\tau)A^\dagger C^{-1}(\tau),
\end{align}
we turn $W^q$ into a $*$-unitary representation. This suggests that if one deforms the inner product in $\mathcal{H}$, we obtain a new Hilbert space $\mathcal{H}_q$ in such a way that the $*$ operation is the Hilbert space adjoint for the new inner product; then, the $q$-symmetric subspace of $\mathcal{H}$ will just be the symmetric subspace of $\mathcal{H}_q$, that is, $\text{Sym}_q \, \mathcal{H}=\text{Sym} \, \mathcal{H}_q$.

Indeed, consider the following bilinear form in $\mathcal{H}$:
\begin{align}
    Q(\psi,\phi)=\braket{\psi|C^{-1}(\tau)|\phi}.
\end{align}
From the properties of $C(\tau)$, it is clear that this form is strictly positive for  {$q>0$}. Therefore, it can be used to define a new inner product, denoted by $\braket{\psi|\phi}_q=Q(\psi,\phi)$.

It is also immediate to check that 
\begin{align}
    Q(\psi,A\phi)=Q(A^*\psi,\phi),
\end{align}
so, in the new inner product, the $*$ operation is the Hilbert space adjoint. Remarkably, the deformation is local for each qubit, by which we mean that
\begin{align}
    \mathcal{H}_q=\otimes \mathcal{H}_{i,q},
\end{align}
where the inner product in $\mathcal{H}_{i,q}$ is given by
\begin{align}
\braket{\psi_i|\phi_i}_q=\braket{\psi_i|q^{\frac12 (N+1-2i)J^3}|\phi_i}.
\label{eq:prodq}
\end{align}

As an application of this fact, one can analyze, for example, the projector onto the $q$-symmetric subspace $\mathcal{H}$ in terms of the projector onto the (undeformed) symmetric subspace of $\mathcal{H}_q$.  Let $\pi_q$ be the projection onto the symmetric subspace of $\mathcal{H}_q$, which takes the form 
\begin{align}
    \pi_q=\frac{1}{N!}\sum_{\sigma \in S_N}W^{q}(\sigma).
    \label{eq:piq}
\end{align}
 {Since $\pi_q$ is the projection onto the $q$-symmetric subspace}, it is clear that $\text{Im}\pi_q=\text{Sym}\mathcal{H}_q=\text{Sym}_q\mathcal{H}$, and also that $\pi_q^2=\pi_q$ and $\pi_q^{*}=\pi_q$. However, $\pi_q$ is not an orthogonal projection with respect to the old inner product, because for that product it is not a self-adjoint operator.  Indeed,
\begin{align}
    \text{Ker}\;\pi_q=(\text{Sym}_q\mathcal{H})^{\perp_q},
\end{align}
where $\perp_q$ denotes the orthogonal complement for the new product, which does not coincide with that of the old product. Actually, the adjoint of $\pi_q$ fulfills
\begin{align}
   ( \pi_q)^{\dagger}=\frac{1}{N!}\sum_{\sigma \in S_N}(W^{q}(\sigma))^{\dagger}=\frac{1}{N!}\sum_{\sigma \in S_N}W^{q^{-1}}(\sigma^{-1})=\pi_{q^{-1}}.
\end{align}

\begin{example}
\label{ex:BFN2}
Following with Example \ref{ex:WqN2} where we computed the only $q$-transposition for the case $N=2$ qubits and taking into account that the order of $S_2$ is $2$, we have that 
\begin{equation}
    \pi_q = \frac{1}{2} (\mathrm{Id}_{4\times 4} + W_1^q) = 
\frac{1}{2} 
\left(
\begin{array}{cccc}
 2 & 0 & 0 & 0 \\
 0 & 1 & q^{-\frac{1}{2}}& 0 \\
 0 & q^{\frac{1}{2}} & 1 & 0 \\
 0 & 0 & 0 & 2 \\
\end{array}
\right) .
\end{equation}
From this expression it can be directly seen that $\mathrm{Im} \, \pi_q$ is generated by the three $q$-Dicke states given in Example \ref{ex:WqN2} and thus $\mathrm{Im} \, \pi_q \approx \mathrm{Sym}_q\mathcal{H}$. Moreover, 
\begin{equation}
    \mathrm{Ker} \, \pi_q = \mathrm{span} \left\{ \frac{1}{\sqrt{ {[2]_q}}} \begin{pmatrix}
        0 \\
        - q^{-\frac{1}{4}} \\
        q^{\frac{1}{4}} \\
        0 \\
    \end{pmatrix}
    \right\} .
\end{equation}
Finally, the matrix associated to the product \eqref{eq:prodq} is given by
\begin{equation}
Q=\left(
\begin{array}{cccc}
 1 & 0 & 0 & 0 \\
 0 & q^{\frac{1}{2}} & 0 & 0 \\
 0 & 0 & q^{-\frac{1}{2}} & 0 \\
 0 & 0 & 0 & 1 \\
\end{array}
\right)
\end{equation}
and it is immediate to check that $\text{Ker}\;\pi_q \approx (\mathrm{Sym}_q \, \mathcal{H})^{\perp_q}$.

\hfill $\square$
\end{example}

\begin{example}
\label{ex:BFN3}
For $N=3$ qubits, in Example \ref{ex:WqN2} we computed the two $q$-transpositions 
\begin{equation}
    W_1^q = W \begin{pmatrix}
    1 & 2 & 3 \\
    2 & 1 & 3 \\
\end{pmatrix}, \qquad \qquad
    W_2^q = W \begin{pmatrix}
    1 & 2 & 3 \\
    1 & 3 & 2 \\
\end{pmatrix} .
\end{equation}
Since the order of $S_3$ is $3! = 6$, we need to compute the representation associated to the rest of the elements, which are explicitly given by 
\begin{equation}
W(\tau(1)) = W \begin{pmatrix}
    1 & 2 & 3 \\
    3 & 2 & 1 \\
\end{pmatrix} =
\left(
\begin{array}{cccccccc}
 1 & 0 & 0 & 0 & 0 & 0 & 0 & 0 \\
 0 & 0 & 0 & 0 & q^{-1} & 0 & 0 & 0 \\
 0 & 0 & 1 & 0 & 0 & 0 & 0 & 0 \\
 0 & 0 & 0 & 0 & 0 & 0 & q^{-1} & 0 \\
 0 & q & 0 & 0 & 0 & 0 & 0 & 0 \\
 0 & 0 & 0 & 0 & 0 & 1 & 0 & 0 \\
 0 & 0 & 0 & q & 0 & 0 & 0 & 0 \\
 0 & 0 & 0 & 0 & 0 & 0 & 0 & 1 \\
\end{array}
\right) ,
\end{equation}
for the one associated to the order-reversing permutation $\tau(1)$, and
\begin{equation}
    W \begin{pmatrix}
    1 & 2 & 3 \\
    2 & 3 & 1 \\
\end{pmatrix} =
\left(
\begin{array}{cccccccc}
 1 & 0 & 0 & 0 & 0 & 0 & 0 & 0 \\
 0 & 0 & q^{-\frac{1}{2}} & 0 & 0 & 0 & 0 & 0 \\
 0 & 0 & 0 & 0 & q^{-\frac{1}{2}} & 0 & 0 & 0 \\
 0 & 0 & 0 & 0 & 0 & 0 & q^{-1} & 0 \\
 0 & q & 0 & 0 & 0 & 0 & 0 & 0 \\
 0 & 0 & 0 & q^{\frac{1}{2}} & 0 & 0 & 0 & 0 \\
 0 & 0 & 0 & 0 & 0 & q^{\frac{1}{2}} & 0 & 0 \\
 0 & 0 & 0 & 0 & 0 & 0 & 0 & 1 \\
\end{array}
\right) ,
\end{equation}
\begin{equation}
    W \begin{pmatrix}
    1 & 2 & 3 \\
    3 & 1 & 2 \\
\end{pmatrix} =
\left(
\begin{array}{cccccccc}
 1 & 0 & 0 & 0 & 0 & 0 & 0 & 0 \\
 0 & 0 & 0 & 0 & q^{-1} & 0 & 0 & 0 \\
 0 & q^{\frac{1}{2}} & 0 & 0 & 0 & 0 & 0 & 0 \\
 0 & 0 & 0 & 0 & 0 & q^{-\frac{1}{2}} & 0 & 0 \\
 0 & 0 & q^{\frac{1}{2}} & 0 & 0 & 0 & 0 & 0 \\
 0 & 0 & 0 & 0 & 0 & 0 & q^{-\frac{1}{2}} & 0 \\
 0 & 0 & 0 & q & 0 & 0 & 0 & 0 \\
 0 & 0 & 0 & 0 & 0 & 0 & 0 & 1 \\
\end{array}
\right) ,
\end{equation}
for the two remaining cycles of order $3$. Now, from \eqref{eq:piq} we have
\begin{equation}
\begin{split}
\pi_q &= \frac{1}{6} \left(\mathrm{Id}_{8\times 8} + W_1^q + W_2^q + W(\tau(1)) + W \begin{pmatrix}
    1 & 2 & 3 \\
    2 & 3 & 1 \\
\end{pmatrix} + W \begin{pmatrix}
    1 & 2 & 3 \\
    3 & 1 & 2 \\
\end{pmatrix} \right) \\
&\qquad\qquad\qquad=
\frac{1}{3}\left(
\begin{array}{cccccccc}
 3 & 0 & 0 & 0 & 0 & 0 & 0 & 0 \\
 0 & 1 & q^{-\frac{1}{2}} & 0 & q^{-1} & 0 & 0 & 0 \\
 0 & q^{\frac{1}{2}} & 1 & 0 & q^{-\frac{1}{2}} & 0 & 0 & 0 \\
 0 & 0 & 0 & 1 & 0 & q^{-\frac{1}{2}} & q^{-1} & 0 \\
 0 & q & q^{\frac{1}{2}} & 0 & 1 & 0 & 0 & 0 \\
 0 & 0 & 0 & q^{\frac{1}{2}} & 0 & 1 & q^{-\frac{1}{2}} & 0 \\
 0 & 0 & 0 & q & 0 & q^{\frac{1}{2}} & 1 & 0 \\
 0 & 0 & 0 & 0 & 0 & 0 & 0 & 3 \\
\end{array}
\right) .
\end{split}
\end{equation}
It is straightforward to check that $\mathrm{Im} \, \pi_q$ is generated by the 4 $q$-Dicke states given in Example \ref{ex:WqN3} and thus $\mathrm{Im} \, \pi_q \approx \mathrm{Sym}_q\mathcal{H}$. Furthermore, we have 
\begin{equation}
    \mathrm{Ker} \, \pi_q = \mathrm{span} \left\{  
    \frac{1}{\sqrt{q + q^{-1}}} \begin{pmatrix}
        0 \\
        0 \\
        0 \\
        -q^{-\frac{1}{2}} \\
        0 \\
        0 \\
        q^{\frac{1}{2}} \\
        0 \\
    \end{pmatrix} ,
        \frac{1}{\sqrt{q + q^{-1}}} \begin{pmatrix}
        0 \\
        -q^{-\frac{1}{2}} \\
        0 \\
        0 \\
        q^{\frac{1}{2}} \\
        0 \\
        0 \\
        0 \\
    \end{pmatrix} ,
    \frac{1}{\sqrt{ {[2]_q}}} \begin{pmatrix}
        0 \\
        0 \\
        0 \\
        -q^{-\frac{1}{4}} \\
        0 \\
        q^{\frac{1}{4}} \\
        0 \\
        0 \\
    \end{pmatrix} ,
    \frac{1}{\sqrt{ {[2]_q}}} \begin{pmatrix}
        0 \\
        -q^{-\frac{1}{4}} \\
        q^{\frac{1}{4}} \\
        0 \\
        0 \\
        0 \\
        0 \\
        0 \\
    \end{pmatrix}
    \right\} .
\end{equation}
The matrix associated to the product \eqref{eq:prodq} reads
\begin{equation}
    Q = 
\left(
\begin{array}{cccccccc}
 1 & 0 & 0 & 0 & 0 & 0 & 0 & 0 \\
 0 & q & 0 & 0 & 0 & 0 & 0 & 0 \\
 0 & 0 & 1 & 0 & 0 & 0 & 0 & 0 \\
 0 & 0 & 0 & q & 0 & 0 & 0 & 0 \\
 0 & 0 & 0 & 0 & q^{-1} & 0 & 0 & 0 \\
 0 & 0 & 0 & 0 & 0 & 1 & 0 & 0 \\
 0 & 0 & 0 & 0 & 0 & 0 & q^{-1} & 0 \\
 0 & 0 & 0 & 0 & 0 & 0 & 0 & 1 \\
\end{array}
\right) .
\end{equation}
and it is immediate to check that $\text{Ker}\;\pi_q \approx (\text{Sym}_q\mathcal{H})^{\perp_q}$.

\hfill $\square$
\end{example}


\section{Applications}
\label{sec:apl}
In this section we introduce several potential applications to quantum informational and metrological tasks of $q$-deformed symmetric subspaces of qubit chains. One could conjecture that, given that the deformed symmetric subspace shares many structural properties with the symmetric subspace, one can find applications by mimicking those of symmetric states (see \cite{marconi2025} for an excellent review) and ponder how these can be adapted to the q-deformed versions. In this section we shall not demonstrate exhaustively and rigorously any results, but will hint at the potentiality of $q-$symmetric states to retain the characteristics that make symmetric states useful in the applications.

\subsection{Entanglement of q-symmetric states} 
 In the context of bipartite entanglement, one can completely characterize the entanglement of pure states, but the entanglement of mixed states for high dimensional systems is difficult since there is no single criterion to tell apart entangled from separable states. Only in low dimensional systems, such as qubit pairs, qubits and qutrits, or Gaussian states of continuous variable quantum systems of a mode with a bath, one can have a criterion, namely the positivity of the partial trace, to distinguish entangled and separable states.  
Characterizing multipartite entanglement is even a more complex task, which is difficult even for pure entangled states. It is common to tackle the issue by resorting to the most, say, unambiguous way of characterizing entanglement, which is to consider the distance of a state to the set of separable states. For instance, consider a pure state in an $N$ partite quantum system (which for simplicity of the exposition we consider identical), say $\ket{\Psi}$, one can define the following quantity
\begin{align}
    G[\ket{\Psi}]=\underset{\{\phi_i\}}{\text{max}}|\!\braket{\Psi|\{\phi_i\}}\!|
\end{align}
where 
\begin{align}
    \ket{\{\phi_i\}}=\ket{\phi_1}...\ket{\phi_N}
\end{align}
are product states.
The quantity $G[\ket{\Psi}]$ is called the geometric measure of entanglement. Overall, this quantity can be written as an optimization problem over $k^N$ variables, where $k$ is the base dimension, and subjected to the constraint that the states $\ket{\phi_i}$ are normalized. It is clear then that the complexity of the problem scales exponentially, and also in general one cannot expect to retrieve analytical results from this expression.    

Symmetric states, however, can be shown to drastically reduce the complexity of the problem. For symmetric states, the state achieving the maximum in the optimization problem can be shown to be symmetric \cite{hubener}, which means that one only has to optimize over families of states of the form $\otimes^N\phi$. In the context of qubits, for instance, one has that one has to optimize over the Bloch sphere, which is a problem that is generally tractable analytically.  One can calculate then the geometric measure of entanglement for Dicke states \cite{wei2003geometric}:
\begin{align}
       G[\ket{D^m_N}]=\sqrt{\binom{N}{m}}\left(\frac{m}{N}\right)^{\frac{m}{2}}\left(\frac{N-m}{N}\right)^{\frac{N-m}{2}}.
\end{align}
It would not be surprising if one could also calculate the geometric measure of entanglement for $q$-Dicke states. In our formalism, it is clear that, in the case of two qubits
\begin{align}
    \braket{\Psi|\phi_1\rangle|\phi_2}=Q(\Psi,C\phi_1\otimes\phi_2)=Q(\Psi, q^{-\frac{J^3_1}{2}}\phi_1\otimes q^{\frac{J^3_{2}}{2}}\phi_2).
\end{align}
Now, because, $Q$ is an inner product and $\Psi$ is q-symmetric, the form 
\begin{align}
    \mathcal{Q}(\bar\phi,\bar\phi')=\braket{\Psi|\bar\phi\rangle|\bar\phi'}
\end{align}
is symmetric and positive, from which one gets that the maximum is given when $\bar\phi=\bar\phi'=\nu$, where $\nu$ is the maximum eigenvalue of the form $\mathcal{Q}$. These leads to the conclusion that the maximum for two qubits is given by
vectors such that $\phi_2= q^{-J^3_1}\phi_1$. It turns out that such vectors are $q$-symmetric, as it is easy to check. Therefore, the closest separable state to a $q$-symmetric state for two qubits is also $q$-symmetric. If we conjecture that a similar relation holds for arbitrary $q$-Dicke states, it would mean that we can construct whole new families of states whose multipartite entanglement has an analytic expression. Further, numerical evidence shows that the entanglement of these deformed states is generally degraded with respect to the undeformed case, and this would mean that these states are less costly in terms of resources while retaining a lot of desirable structure and may be better suited for some applications.

\subsection{Metrology}
In the context of quantum metrology, symmetric states are important since they saturate the optimal parameter estimation of certain quantities such as the magnetization and the energy of spin chains \cite{oszmaniek2016}. Indeed, for a 1-parameter family of states of the form $\ket{\phi}_\theta=e^{-i A\theta}\phi$, it can be shown that the optimal random variable $\bar \theta$ constructed with measurements over $\ket{\phi}_\theta$ satisfies
\begin{align}
 \big \langle(  \bar{\theta}-\braket{\bar\theta})\big \rangle^2 \big \rangle \big \langle(  A-\openone\braket{A}_\phi)^2\big \rangle_\psi >\frac12.
\end{align}
This inequality is known as the quantum Cramer-Rao bound \cite{paris}. Therefore, states that have higher variance are more sensitive to the value of the parameter $\theta$, and allow for a higher estimation. In the context of spin chains, given the (undeformed) Casimir:
\begin{align}
    \mathcal{C}=\Delta_{\frac12}^{(N)}(J^2_x)+\Delta_{\frac12}^{(N)}(J^2_y)+\Delta_{\frac12}^{(N)}(J^2_z)\leq\frac{N}{2}\left(\frac{N}{2}+1\right),
\end{align}
we see that the maximum value of, say $J^2_z$ is given by states such that $\braket{\Delta_{\frac12}^{(N)}(J^2_z)}=\frac{N}{2}(\frac{N}{2}+1)$, which will happen for symmetric states, and similarly for $J_{x,y}$. This shows that the optimal states scale as $N^2$, in contrast with the $N$ scaling expected from separable states, which shows that symmetric entanglement is a resource for metrology in this setting. 

When promoting these properties to the $q$-deformed case, we find

\begin{align}
    \mathcal{C}_q=\Delta_{\frac12}^{(N)}(J^2_x)+\Delta_{\frac12}^{(N)}(J^2_y)+\frac{\sqrt{q}+\frac{1}{\sqrt{q}}}{2}\Delta_{\frac12}^{(N)}([J_z]_q^2)\leq\left[\frac{N}{2}\right]_q\left[\frac{N}{2}+1\right]_q.
\end{align}
We see that there is no direct advantage over the symmetric subspace in estimating parameters involving $J_z$ (In fact, the advantage is the same since the GHZ state also belongs to the $q$-symmetric subspace \cite{marconi2025}). However, the state $\ket{D^{\frac{N}{2}}_N}$, which we could dub the $q$-Werner state, fulfills
\begin{align}
   \braket {D^{\frac{N}{2}}_N|\Delta_{\frac12}^{(N)}(J^2_y)|D^{\frac{N}{2}}_N}=\braket {D^{\frac{N}{2}}_N|\Delta_{\frac12}^{(N)}(J^2_x)|D^{\frac{N}{2}}_N}=\left[\frac{N}{2}\right]_q\left[\frac{N}{2}+1\right]_q .
\end{align}
Now, because of the homomorphism property of the coproduct, $ \braket {D^{\frac{N}{2}}_N|\Delta_{\frac12}^{(N)}(J^2_y)|D^{\frac{N}{2}}_N}$ are the variances of selfadjoint operators, which are given in terms of local operators as
\begin{align}
    \Delta_{\frac12}^{(N)}(J_{x,y})=\frac12\sum_i q^{\sum_{j<i} \frac{\sigma^i_z}{4}-\sum_{j>i} \frac{\sigma^j_{z}}{4}}\sigma^i_{x,y} ,
\end{align}
where we understand $\sigma^i_{x,y,z}=\openone_i\otimes^{(i-1)}...\sigma_{x,y,z}...\otimes\openone$. These operators represent sums of local operators that have a bias depending on the number of excitations left or right of the position $i$. Importantly, we see that introducing this bias improves significantly the maximum sensitivity of states in the Hilbert space $\mathcal{H}$, since the variance grows exponentially as $N$ grows rather than quadratically.

\section{Concluding remarks and open problems}
\label{sec:concl}

This work presents a comprehensive analysis of quantum group deformations of the symmetric subspace in multi-qubit systems.
Our starting point was the duality between the permutation group acting on a system of $N$ qubits and the representations of $\mathfrak{su}(2)$. Specifically, we emphasized that the symmetric subspace carries an irreducible representation of $\mathfrak{su}(2)$. This characterization allowed us to discuss the deformations of the symmetric subspace as deformations of the universal enveloping algebra of $\mathfrak{su}(2)$.  {We focused on the case of deformations that essentially preserve the undeformed representation theory, that is, $q$-deformations in which $q$ is not a root of unity.} The coassociativity of these Hopf algebra deformations has allowed us to characterize the deformations through a suitable basis of states, the so-called Dicke states, which are continuously deformed in a one-to-one way as $q$ varies. 

While the dual algebra (in the Schur-Weyl sense) of the deformed $\mathcal{U}_q(\mathfrak{su}(2))$ can be described in terms of the so-called Hecke algebra, our analysis in terms of Dicke states has allowed us to describe the $q$-symmetric subspace in terms of a non-unitary representation of the permutation group, which implies a significant simplification in its description.  {In particular, we have shown that this representation, for $q>0$, can be unitarized by deforming the inner product of the qubit locally in such a way that one can describe these deformations of the symmetric subspace as a the (undeformed) symmetric subspace of some deformed Hilbert space.}

{Besides the applications sketched in the previous section}, future works will address three avenues regarding these deformations. One obvious one is to apply the full generality of the Schur-Weyl duality for base Hilbert spaces of arbitrary dimension, such as multiqutrit systems, which require an analysis on similar terms of {the corresponding $q$-deformation of $\mathfrak{su}(n)$, whose $q$-Dicke states have been introduced in~\cite{Raveh_2024}}. Another obvious generalization is to consider $q$-deformations of subspaces of exchangeable states, describing symmetric density matrices. These two would be indispensable ingredients to describe a deformed version of de Finetti's theorem  \cite{watrous2018theory,Konig2005,Gross}, for which one would aim to describe the local features of such states for general base dimension. Finally, there is the case which $q$ is a root of unity,  {which we have not addressed here}, where the representation theory is not mapped one-to-one with the undeformed one, and significantly different results could arise. Work on all these lines is in progress.

\section*{Acknowledgements}

The authors acknowledge partial support from the grant PID2023-148373NB-I00 funded by MCIN/AEI/ 10.13039/501100011033/FEDER -- UE, and the Q-CAYLE Project funded by the Regional Government of Castilla y León (Junta de Castilla y León) and the Ministry of Science and Innovation MICIN through NextGenerationEU (PRTR C17.I1). I. Gutierrez-Sagredo thanks Universidad de La Laguna, where part of the work has been done, for the hospitality and support.

\end{document}